\documentclass[12pt]{article}
\usepackage{amssymb}
\usepackage{amsmath}

\makeatletter
\def\numberbysection{\@addtoreset{equation}{section}
 	\def\theequation{\thesection.\arabic{equation}}}
\makeatother

\numberbysection

\newcommand{\be}{\begin{eqnarray}}
\newcommand{\ee}{\end{eqnarray}}
\newcommand{\non}{\nonumber}
\newcommand{\tr}{\mathop{\rm tr}\nolimits}
\newcommand{\kk}{\kappa}
\newcommand{\id}{\mathbb{I}}

\begin{document}

\begin{titlepage}
\strut\hfill UMTG--239
\vspace{.5in}
\begin{center}

\LARGE Functional relations and Bethe Ansatz \\
\LARGE for the XXZ chain\\[1.0in]
\large Rafael I. Nepomechie\\[0.8in]
\large Physics Department, P.O. Box 248046, University of Miami\\[0.2in]  
\large Coral Gables, FL 33124 USA\\

\end{center}

\vspace{.5in}

\begin{abstract}
There is an approach due to Bazhanov and Reshetikhin for solving
integrable RSOS models which consists of solving the functional
relations which result from the truncation of the fusion hierarchy. 
We demonstrate that this is also an effective means of solving
integrable vertex models.  Indeed, we use this method to recover the
known Bethe Ansatz solutions of both the closed and open XXZ quantum
spin chains with $U(1)$ symmetry.  Moreover, since this method does
not rely on the existence of a pseudovacuum state, we also use this
method to solve a special case of the open XXZ chain with nondiagonal
boundary terms.
\end{abstract}
\end{titlepage}

\setcounter{footnote}{0}

\section{Introduction}\label{sec:intro}

There are several well-known methods of deriving the Bethe Ansatz (BA)
solution of integrable vertex models: the coordinate BA \cite{Be, Ba1,
ABBBQ}, the $T-Q$ approach \cite{Ba1}, the algebraic BA \cite{FT, KS1,
Sk}, the analytic BA \cite{Re}, and the functional BA \cite{Sk2}.  We
present here yet another method, which entails solving the functional
relations which result from the truncation of a model's fusion
hierarchy.  This approach was (to our knowledge) first developed for
RSOS models \cite{ABF} by Bazhanov and Reshetikhin in \cite{BR}, but
until now has not been applied to vertex-type models.  An important
feature of this method is that, unlike some of the other approaches,
it does not rely on the existence of a pseudovacuum (reference) state.

Our primary motivation for this work comes from the long outstanding
problem of solving the open spin-${1\over 2}$ XXZ quantum spin chain
with nondiagonal boundary terms, defined by the Hamiltonian \cite{dVGR}
\be
{\cal H }&=& {1\over 2}\Big\{ \sum_{n=1}^{N-1}\left( 
\sigma_{n}^{x}\sigma_{n+1}^{x}+\sigma_{n}^{y}\sigma_{n+1}^{y}
+\cosh \eta\ \sigma_{n}^{z}\sigma_{n+1}^{z}\right)\non \\
&+&\sinh \eta \Big( \coth \xi_{-} \sigma_{1}^{z}
+ {2 \kk_{-}\over \sinh \xi_{-}}\sigma_{1}^{x} 
- \coth \xi_{+} \sigma_{N}^{z}
- {2 \kk_{+}\over \sinh \xi_{+}}\sigma_{N}^{x} \Big) \Big\} \,,
\label{Hamiltonian}
\ee
where $\sigma^{x} \,, \sigma^{y} \,, \sigma^{z}$ are the standard
Pauli matrices, $\eta$ is the bulk anisotropy parameter, $\xi_{\pm} \,,
\kk_{\pm}$ are arbitrary boundary parameters, and $N$ is the number of
spins.  This model is integrable. Indeed, the Hamiltonian is obtained 
from the commuting transfer matrix \cite{Sk} constructed with
the nondiagonal $K$ matrix found in \cite{dVGR, GZ} together with
the standard XXZ $R$ matrix. 

Solving this problem (e.g., determining the energy eigenvalues in
terms of roots of a system of Bethe Ansatz equations) is a crucial
step in formulating the thermodynamics of the spin chain and of the
boundary sine-Gordon model \cite{GZ}.  Moreover, this problem has
important applications in condensed matter physics and statistical
mechanics.

A fundamental difficulty is that, in contrast to the special case of
diagonal boundary terms (i.e., $\kk_{\pm}=0$, in which case ${\cal H
}$ has a $U(1)$ symmetry) considered in \cite{ABBBQ, Sk}, a simple
pseudovacuum state does {\it not} exist (e.g., the state with all
spins up is not an eigenstate of the Hamiltonian).  

Some progress on this problem was made recently in \cite{XX, XXZ}. 
Namely, for bulk anisotropy value $\eta = {i \pi\over p+1}\,, \qquad
p= 1 \,, 2 \,, \ldots \,,$ (and hence $q \equiv e^{\eta}$ is a root of
unity, satisfying $q^{p+1}=-1$), an exact $(p+1)$-order functional
relation for the fundamental transfer matrix was proposed.  The key
observation is that the fused spin-${p+1\over 2}$ transfer matrix can
be expressed in terms of a lower-spin transfer matrix, resulting in
the truncation of the fusion hierarchy.  \footnote{This is distinct
from the observation due to Belavin {\it et al.} \cite{BS, BGF} that,
for the special case of quantum-group symmetry (i.e., $\kk_{\pm}=0$,
$\xi_{\pm}\rightarrow \infty$), the fused transfer matrix $t^{({p\over
2})}(u)$ vanishes after quantum group reduction.} The simplest case
$p=1$, which corresponds to the XX chain, is analyzed in \cite{XX}. 

Although sets of equations for the eigenvalues of the transfer matrix
were found in \cite{XXZ}, these equations do not have the standard
Bethe Ansatz form, and they become increasingly complicated as the
value of $p$ increases.  Moreover, one would like to solve the model
for general values of $\eta$, i.e., not just for the discrete values
corresponding to roots of unity.

We also achieve here some progress on these questions.  In
particular, from the functional relations in \cite{XXZ}, we obtain
standard Bethe Ansatz equations for the transfer matrix eigenvalues
for general values of $\eta$, albeit only for the special case
\be
\kk_{+} = \kk_{-} \equiv \kk \ne 0 \,, \qquad 
\xi_{+} = \xi_{-} \equiv \xi \,, \qquad N =  \mbox{odd} \,.
\label{specialcase}
\ee
Unfortunately, we have not yet succeeded to obtain corresponding
results for general values of the boundary parameters.

The outline of this article is as follows.  In Section
\ref{sec:closed}, we consider as a warm-up the case of the closed XXZ
chain, and provide a new derivation of the well-known Bethe Ansatz
solution.  In Section \ref{sec:open}, we turn to the open XXZ chain. 
Using the functional relations proposed in \cite{XXZ}, we first
recover the solution of Alcaraz {\it et al.} \cite{ABBBQ} and Sklyanin
\cite{Sk} for the diagonal case $\kk_{\pm} = 0$, and we then give the
solution for the nondiagonal case (\ref{specialcase}).  We conclude
with a brief discussion of our results in Section \ref{sec:discuss}.

\section{The closed chain}\label{sec:closed}

The closed (i.e., with periodic boundary conditions) spin-${1\over 2}$
XXZ quantum spin chain is defined by the Hamiltonian
\be
H = {1\over 2} \sum_{n=1}^{N}\left( 
\sigma_{n}^{x}\sigma_{n+1}^{x}+\sigma_{n}^{y}\sigma_{n+1}^{y}
+ \cosh \eta\ \sigma_{n}^{z}\sigma_{n+1}^{z}\right) \,,
\label{closedHamiltonian}
\ee
where $\vec \sigma_{N+1} = \vec \sigma_{1}$.  As noted in the
Introduction, there are various methods of deriving the Bethe Ansatz
solution of this model.  As a warm-up for the open-chain problem, we
now give another derivation of this solution, which involves
solving the model's functional relations.  In Sec. 
\ref{subsec:closedfuncreltns} we derive the functional relations, and
then in Sec.  \ref{subsec:closedBA} we proceed to solve them.

\subsection{Functional relations}\label{subsec:closedfuncreltns}

In this subsection, we begin by briefly reviewing the construction of
the (fused) transfer matrices of the closed XXZ chain.  We then recall
the so-called fusion hierarchy which these transfer matrices obey. 
Finally, we give an identity which truncates the fusion hierarchy and
leads to the desired functional relations.

The fundamental spin-$({1\over 2},{1\over 2})$ XXZ $R$ matrix is given
by the $4 \times 4$ matrix
\be
R(u) = \left( \begin{array}{cccc}
	\sinh  (u + \eta) &0            &0           &0            \\
        0                 &\sinh  u     &\sinh \eta  &0            \\
	0                 &\sinh \eta   &\sinh  u    &0            \\
	0                 &0            &0           &\sinh  (u + \eta)
\end{array} \right) \,,
\label{bulkRmatrix}
\ee 
where $\eta$ is the anisotropy parameter.  It is a solution of the
Yang-Baxter equation
\be
R_{12}(u-v)\ R_{13}(u)\ R_{23}(v)
= R_{23}(v)\ R_{13}(u)\ R_{12}(u-v) \,.
\label{YB}
\ee 
(See, e.g., \cite{KS2, KBI}.)  The fused spin-$(j,{1\over
2})$ $R$ matrix ($j = {1\over 2} \,, 1 \,, {3\over 2}\,, \ldots $) is
given by \cite{KS2, KRS}
\be
R_{\langle 1 \ldots 2j \rangle 2j+1}(u) = P^{+}_{1 \ldots 2j} 
R_{1, 2j+1}(u) R_{2, 2j+1}(u+\eta) \ldots R_{2j, 2j+1}(u +(2j-1)\eta) 
P^{+}_{1 \ldots 2j} \,.
\label{fusedR}
\ee 
The (undeformed) projectors are defined by
\be
P^{\pm}_{1 \ldots m}={1\over m!}\sum_{\sigma}(\pm 1)^{\sigma} {\cal 
P}_{\sigma} \,,
\ee
where the sum is over all permutations $\sigma = (\sigma_{1} \,, 
\ldots \sigma_{m})$ of $(1\,, \ldots \,, m)$, and ${\cal P}_{\sigma}$ 
is the permutation operator in the space $\otimes_{k=1}^{m} {\cal 
C}^{2}$. For instance,
\be
P^{+}_{12} &=& {1\over 2}(\id + {\cal P}_{12}) \,, \non \\
P^{+}_{123} &=& {1\over 6}(\id + {\cal P}_{23}{\cal P}_{12}+{\cal 
P}_{12}{\cal P}_{23}+{\cal P}_{12}+{\cal P}_{23}+{\cal P}_{13}) \,,
\ee 
where $\id$ is the identity matrix.

The closed-chain transfer matrix $t^{(j)}(u)$, which is
constructed using a spin-$j$ auxiliary space, is defined by
\be
t^{(j)}(u)= \tr_{1 \ldots 2j} 
T_{\langle 1 \ldots 2j \rangle}(u)  \,,
\label{closedfusedtransfer}
\ee 
where the fused monodromy matrix is defined by
\be
T_{\langle 1 \ldots 2j \rangle}(u) 
= R_{\langle 1 \ldots 2j \rangle N}(u) \ldots 
R_{\langle 1 \ldots 2j \rangle 1}(u)  \,,
\ee 
and $N$ corresponds to the number of spins of the chain.
One can show that
\be
T_{\langle 1 \ldots 2j \rangle}(u) 
= P^{+}_{1 \ldots 2j} 
T_{1}(u) T_{2}(u+\eta) \ldots T_{2j}(u +(2j-1)\eta) 
P^{+}_{1 \ldots 2j}  \,.
\ee 
These transfer matrices constitute commutative families
\be
\left[ t^{(j)}(u)\,, t^{(k)}(v) \right] = 0  \,.
\label{commutativity}
\ee 
The fundamental transfer matrix $t(u) \equiv t^{({1\over 2})}(u)$
contains the Hamiltonian (\ref{closedHamiltonian}),
\be
H \propto {\partial\over \partial u}\log t(u) 
\Big\vert_{u=0} + const. \,,
\ee
and has the periodicity property
\be
t(u+ i\pi) = (-1)^{N}\ t(u) \,.
\label{closedperiodicity}
\ee 

The fusion hierarchy for the XXZ chain is given by \cite{KS2, KiR}
\be
t^{(j)}(u)\ t^{({1\over 2})}(u+2j\eta) = 
\delta(u+(2j-1)\eta)\ t^{(j-{1\over 2})}(u) 
+ t^{(j+{1\over 2})}(u) \,, \qquad 
j = {1\over 2} \,, 1 \,, {3\over 2}\,, \ldots \,,
\label{closedhierarchy}
\ee
where $t^{(0)}(u) = \id$, and the quantum
determinant \cite{KS2, IK} $\delta(u)$ is given by
\be
\delta(u) = \tr_{12} P^{-}_{12}\ T_{1}(u) T_{2}(u+\eta) 
= \left(-\zeta(u+\eta) \right)^{N} \,,
\ee
where
\be
\zeta(u) = -\sinh (u+\eta) \sinh(u-\eta) \,.
\label{zeta}
\ee 

The key fact in deriving the functional relations is that
for anisotropy values
\be
\eta = {i \pi\over p+1}\,, \qquad p= 1 \,, 2 \,, \ldots \,,
\label{etavalues}
\ee
the fused transfer matrices satisfy the identity
\be
t^{({p+1\over 2})}(u) = (-1)^{N} \delta(u-\eta) 
\left[ t^{({p-1\over 2})}(u+ \eta) + 
(1+ (-1)^{N}) \nu(u)^{N} F \right] \,, 
\label{closedtruncation}
\ee
where 
\be
\nu(u) = - {1\over \zeta(u)}\prod_{k=0}^{p} \sinh(u + k \eta) 
= - {1\over \zeta(u)} \left({i\over 2}\right)^{p} \sinh((p+1)u) \,,
\label{nu}
\ee
and 
\be
F=\prod_{k=1}^{N} \sigma^{z}_{k} \,.
\label{Foperator}
\ee
The remarkable result (\ref{closedtruncation}), to which we refer as 
the ``truncation identity,'' follows directly from
Eq.  (4.13) in Ref.  \cite{XXZ}, which relies on the quantum-group
construction \cite{KuR} of higher-spin $R$ matrices.

The fact that the spin-${p+1\over 2}$ transfer matrix can be expressed
in terms of a lower-spin transfer matrix leads to the truncation of
the fusion hierarchy, which in turn leads to a $(p+1)$-order
functional relation for the fundamental transfer matrix.  For
instance, for the case $p=2$, Eqs.  (\ref{closedhierarchy}) and
(\ref{closedtruncation}) lead to the third-order functional relation
\be
t(u) t(u+\eta) t(u+2\eta) - \delta(u) t(u+2\eta) 
- \delta(u+\eta) t(u) \qquad \qquad \qquad \non \\
- (-1)^{N} \delta(u-\eta) t(u+\eta) 
- (1+ (-1)^{N}) \delta(u-\eta) \nu(u)^{N} F =0 \,.
\ee
Similar higher-order functional relations have been obtained for RSOS
models \cite{Ba1, BP, BR} and for the 8-vertex model \cite{Ba2}.  We
emphasize that, contrary to the commonly-held misconception (see, e.g.
\cite{KNS}), the fusion hierarchies of vertex models {\it do}
truncate, for the $\eta$ values (\ref{etavalues}).

The commutativity relation (\ref{commutativity}) with $j=k={1\over 
2}$ and the fact $[ F \,, t(u) ]=0$ imply that $t(u)$ and $F$ can be 
simultaneously diagonalized,
\be
t(u) |\Lambda^{(\pm 1)} \rangle &=& \Lambda^{(\pm 1)}(u) 
| \Lambda^{(\pm 1)} \rangle \,, \non \\
F |\Lambda^{(\pm 1)} \rangle &=& \pm | \Lambda^{(\pm 1)} \rangle \,,
\label{eigenvalueproblem}
\ee
where the eigenstates $| \Lambda^{(\pm 1)} \rangle$ are independent of
$u$.  Acting on these eigenstates with the functional relations, one
obtains the corresponding relations for the eigenvalues.

\subsection{Bethe Ansatz solution}\label{subsec:closedBA}

We now proceed to solve the functional relations for the eigenvalues
of the fundamental transfer matrix.  Following Ref.  \cite{BR}, we
observe that the functional relations for $p \ge 2$ can be represented
in a compact form as the determinant of a $(p+1) \times (p+1)$ matrix:
\be
\det \left(
\begin{array}{cccccccc}
    \Lambda^{(F)}_{0} & -h_{-1} & 0 & 0 & \ldots & 0 & 0 & -F h_{0}  \\
    -h_{1} & \Lambda^{(F)}_{1} & -h_{0} & 0 & \ldots & 0 & 0 & 0  \\
    0 & -h_{2} & \Lambda^{(F)}_{2} & -h_{1} & \ldots & 0 & 0 & 0  \\
     &  &  & & \ddots    \\
    0 & 0 & 0 & 0 & \ldots & -h_{p-1} & \Lambda^{(F)}_{p-1} & -h_{p-2}  \\
    -F h_{p-1} & 0 & 0 & 0 & \ldots & 0 & -h_{p} & \Lambda^{(F)}_{p}
\end{array} \right) = 0 \,,
\label{detform}
\ee
where 
\be
h(u) = \sinh^{N}(u+\eta) \,,
\label{functionf}
\ee
$h_{k} = h(u + \eta k)$, $\Lambda^{(F)}_{k} = \Lambda^{(F)}(u + \eta
k)$, and $F = \pm 1$ now denotes the eigenvalue of the operator in
(\ref{Foperator}).  Let $(Q_{0} \,, Q_{1} \,, \ldots \,, Q_{p})$ be
the null vector of the matrix in (\ref{detform}); i.e.,
\be
\Lambda^{(F)}_{0} Q_{0} - h_{-1} Q_{1} - F h_{0} Q_{p} &=& 0 \,,  
\non \\
-h_{k} Q_{k-1} + \Lambda^{(F)}_{k} Q_{k} - h_{k-1} Q_{k+1} &=& 0 \,, \qquad
k = 1\,, \ldots \,, p-1  \,, \non \\
-F h_{p-1} Q_{0} - h_{p} Q_{p-1} + \Lambda^{(F)}_{p} Q_{p}  &=& 0 \,.
\label{null}
\ee
We make the Ansatz $Q_{k} = Q(u + \eta k)$, where
\be
Q(u) = \prod_{j=1}^{M} \sinh(u - u_{j}) \,,
\label{closedQ}
\ee
for some integer $M$.  Eqs.  (\ref{null}) imply (using $Q_{k} =
(-1)^{M} Q_{k+p+1}$) that the eigenvalues are given by
\be
\Lambda^{(F)}(u) = h(u) {Q(u-\eta)\over Q(u)} 
+ h(u-\eta) {Q(u+\eta)\over Q(u)}  \,, 
\label{closedeigenvalues} 
\ee
and 
\be 
F = (-1)^{M} \,. 
\label{FMrelation}
\ee
We verify that the result (\ref{closedeigenvalues}) is consistent with
the periodicity condition (\ref{closedperiodicity}).
The requirement that $\Lambda^{(F)}(u)$ be analytic at $u=u_{j}$ 
yields the Bethe Ansatz equations
\be
{h(u_{j})\over h(u_{j}-\eta)} =  
-{Q(u_{j}+\eta)\over Q(u_{j}-\eta)} \,, \qquad j = 1 \,, \ldots \,, M \,.
\label{closedBAeqs}
\ee

We recognize Eqs.  (\ref{functionf}), (\ref{closedQ}),
(\ref{closedeigenvalues}) and (\ref{closedBAeqs}) as the familiar
Bethe Ansatz result for the eigenvalues of the transfer matrix of the
closed XXZ chain.  Although we have assumed that $\eta$ has the values
(\ref{etavalues}), these results are known to be true for general
values of $\eta$.  Note also that the approach that we have followed
here does not explicitly rely on the existence of a pseudovacuum
state.

A more thorough analysis would also include the diagonalization of
$S^{z}={1\over 2}\sum_{k=1}^{N} \sigma^{z}_{k}$ along with $t(u)$ and
$F$.  By considering the asymptotic behavior of $t(u)$ for $u
\rightarrow \infty$, one can establish that the value of $M$ is
related to the $S^{z}$ eigenvalue, namely, $M={N\over 2}-S^{z}$. 
Since these matters are already well-understood, and since the
open-chain problem (\ref{Hamiltonian}) lacks this additional $U(1)$
symmetry, we do not pursue this issue further.

Finally, we point out that the result (\ref{FMrelation}), which
perhaps is less familiar, can nevertheless be readily obtained within
the algebraic Bethe Ansatz approach \cite{KS1}.  Indeed, as is well
known, the eigenstates of the transfer matrix are given by
\be
| u_{1}\,, \ldots \,, u_{M} \rangle 
= B(u_{1}) \ldots B(u_{M}) | \Omega \rangle \,,
\label{eigenstate}
\ee
where $B(u)$ is a certain creation-like operator, and $| \Omega
\rangle$ is the pseudovacuum state with all spins up.  It is easy to
show that $\{ F \,, B(u) \}=0$ and $F | \Omega \rangle = | \Omega
\rangle$.  Hence, the state $| u_{1}\,, \ldots \,, u_{M} \rangle$
has the eigenvalue $F = (-1)^{M}$.

\section{The open chain}\label{sec:open}

We turn now to the open chain (\ref{Hamiltonian}), which is our main
concern.  Our strategy is to try to generalize the analysis of
the preceding section.  Hence, in Sec.  \ref{subsec:openfuncreltns}
we review the functional relations, and then in Sec. 
\ref{subsec:openBA} we attempt to solve them.  

\subsection{Functional relations}\label{subsec:openfuncreltns}

The fundamental spin-${1\over 2}$ XXZ $K^{-}$ matrix is given
by the $2 \times 2$ matrix \cite{dVGR, GZ}
\be
K^{-}(u) = \left( \begin{array}{cc}
\sinh(\xi_{-} + u)   & \kk_{-} \sinh  2u \\
\kk_{-} \sinh  2u     & \sinh(\xi_{-} - u) 
\end{array} \right) \,,
\label{Kminusmatrix}
\ee 
which evidently depends on two boundary parameters $\xi_{-} \,, 
\kk_{-}$. It is a solution of the boundary Yang-Baxter equation 
\cite{Ch}
\be
R_{12}(u-v)\ K^{-}_{1}(u)\ R_{21}(u+v)\ K^{-}_{2}(v)
= K^{-}_{2}(v)\ R_{12}(u+v)\ K^{-}_{1}(u)\ R_{21}(u-v) \,.
\label{boundaryYB}
\ee 

The fundamental open-chain transfer matrix $t(u)$ is constructed,
following Sklyanin's recipe \cite{Sk}, from the matrix $R(u)$
(\ref{bulkRmatrix}), the matrix $K^{-}(u)$ (\ref{Kminusmatrix}), and
the matrix $K^{+}(u)$ which is equal to $K^{-}(-u-\eta)$ with
$(\xi_{-} \,, \kk_{-})$ replaced by $(\xi_{+} \,, \kk_{+})$.
\footnote{Further details about the construction of this transfer
matrix can be found in \cite{XXZ}.} The Hamiltonian
(\ref{Hamiltonian}) is related to the first derivative of the transfer
matrix,
\be
{\cal H} = {1\over 4 \sinh \xi_{-} \sinh \xi_{+} \sinh^{2N-1} \eta 
\cosh \eta} {\partial t(u) \over \partial u}\Big\vert_{u=0} 
- {\sinh^{2}\eta  + N \cosh^{2}\eta\over 2 \cosh \eta} 
\id \,.
\ee
The transfer matrix has the periodicity property
\be
t(u + i \pi)= t(u) \,,
\label{openperiodicity}
\ee
as well as crossing symmetry
\be
t(-u - \eta)= t(u) \,,
\label{transfercrossing}
\ee
and the asymptotic behavior (for $\kk_{\pm} \ne 0$)
\be
t(u) \sim -\kk_{-}\kk_{+} {e^{u(2N+4)+\eta (N+2)}\over 2^{2N+1}} \id + 
\ldots \qquad \mbox{for} \qquad
u\rightarrow \infty \,.
\label{transfasympt}
\ee

Functional relations for the open XXZ chain (\ref{Hamiltonian}) have
been proposed in Ref.  \cite{XXZ}. 
These relations, which follow from the fusion hierarchy \cite{MN1, Zh}
together with the truncation identity for the $\eta$ values 
(\ref{etavalues}), are given by 
\be
\lefteqn{\Lambda(u) \Lambda(u +\eta) \ldots \Lambda(u + p \eta)} \non \\
&-& \delta (u-\eta) \Lambda(u +\eta) \Lambda(u +2\eta) 
\ldots \Lambda(u + (p-1)\eta) \non \\
&-& \delta (u) \Lambda(u +2\eta) \Lambda(u +3\eta)
\ldots \Lambda(u + p \eta) \non \\
&-& \delta (u+\eta) \Lambda(u) \Lambda(u +3\eta) \Lambda(u +4\eta) 
\ldots \Lambda(u + p \eta) \non \\
&-& \delta (u+2\eta) \Lambda(u) \Lambda(u +\eta) \Lambda(u +4\eta) 
\ldots \Lambda(u + p \eta) - \ldots \non \\
&-& \delta (u+(p-1)\eta) \Lambda(u) \Lambda(u +\eta) 
\ldots \Lambda(u +  (p-2)\eta) \non \\
&+& \ldots  = f(u) \,,
\label{funcrltn}
\ee 
where $\Lambda(u)$ is the eigenvalue of the fundamental open-chain
transfer matrix $t(u)$.  Furthermore, the function $\delta(u)$ is now
defined by
\be
\delta(u) = {\Delta (u)\over \zeta(2u+2\eta)} \,,
\ee
where the quantum determinant $\Delta (u)$ is given by
\be
\Delta (u) &=&
 -\left[\sinh(u + \eta + \xi_{-}) \sinh(u + \eta - \xi_{-})
+ \kk_{-}^{2} \sinh^{2}(2u+2\eta) \right] \non \\
&\times& \left[\sinh(u + \eta + \xi_{+}) \sinh(u + \eta - \xi_{+})
+ \kk_{+}^{2} \sinh^{2}(2u+2\eta) \right] \non \\
&\times& \sinh 2u \sinh(2u+4\eta)\ \zeta(u + \eta)^{2N} \,,
\label{qdeterminant}
\ee
and $\zeta(u)$ is defined in Eq. (\ref{zeta}). Moreover, the 
function $f(u)$ is given by \footnote{In terms of the three functions
$f_{0}$, $f_{1}$, $f_{3}$ used in \cite{XXZ}, the functions
$\delta(u)$ and $f(u)$ are given by 
\be
\delta(u) = {f_{1}(u+\eta)\over f_{0}(u)} \,, \qquad 
f(u) = {f_{3}(u) \over f_{0}(u)} \,. \non 
\ee}
\be
f(u) &=& {(-1)^{p (N+1)}\over 2^{2p(N+1)}} \sinh^{2N}((p+1)u) 
{\cosh^{2}((p+1)u +{i \pi\over 2}\epsilon)\over \cosh^{2}((p+1)u)}
\non \\
&\times& \Big\{ 
n(u \,; \xi_{-} \,, \kk_{-})\ n(u \,; -\xi_{+} \,, \kk_{+}) +
n(u \,; -\xi_{-} \,, \kk_{-})\ n(u \,; \xi_{+} \,, \kk_{+}) \non \\
&\quad&+ 2 (-1)^{N} (-\kk_{-} \kk_{+})^{p+1} \sinh^{2}(2(p+1)u) \Big\}
\,,
\ee
where $\epsilon= 2 \mbox{frac}(p/2)$ equals 0 if $p$ is even, and
equals 1 if $p$ is odd; and the function $n(u \,; \xi \,, \kk)$ is
defined by
\be
n(u \,; \xi \,, \kk) = \sinh \left( (p+1)(\xi +u) \right)  
+ \sum_{l=1}^{\left[{p+1\over 2}\right]}c_{p\,, l}\ 
\kk^{2l} \sinh \left( (p+1)u + (p+1 - 2l) \xi \right) \,,
\label{nfunction}
\ee 
with
\be
c_{p \,, l} = {(p+1)\over l!} \prod_{k=0}^{l-2} (p-l-k) 
\,. \non 
\ee
For instance, for the case $p=3$, the functional relation is given 
by \footnote{The last two terms of the left-hand-side were
accidentally omitted in Eq.  (5.2) of \cite{XXZ}.}
\be
\Lambda(u) \Lambda(u +\eta) \Lambda(u +2\eta) \Lambda(u +3\eta) 
- \delta(u-\eta) \Lambda(u +\eta) \Lambda(u +2\eta) 
\qquad \qquad \qquad \qquad \non  \\
 - \delta(u) \Lambda(u +2\eta) \Lambda(u +3\eta) 
- \delta(u +\eta) \Lambda(u) \Lambda(u +3\eta)   
- \delta(u +2\eta) \Lambda(u) \Lambda(u +\eta)  \non  \\ 
+ \delta(u) \delta(u +2\eta)
+ \delta(u-\eta) \delta(u +\eta) = f(u) \,.
\ee 

\subsection{Bethe Ansatz solution}\label{subsec:openBA}

For general values of the boundary parameters $\kk_{\pm} \,,
\xi_{\pm}$, we have not yet succeeded to find a determinant
representation analogous to (\ref{detform}) of the functional
relations (\ref{funcrltn}).  Nevertheless, for the following two
special cases of the boundary parameters, we have found such
representations.

\subsubsection{The diagonal case $\kk_{\pm}=0$}

The Bethe Ansatz solution for the diagonal case $\kk_{\pm}=0$ is 
already known \cite{ABBBQ, Sk}. Nevertheless, it is instructive to see 
how this solution emerges from the functional relations. Indeed,
when $\kk_{\pm}=0$, the functional relations for $p \ge 2$ 
can be represented as
\be
\det \left(
\begin{array}{cccccccc}
    \Lambda_{0} & -h'_{-1} & 0 & 0 & \ldots & 0 & 0 & -h_{0}  \\
    -h_{1} & \Lambda_{1} & -h'_{0} & 0 & \ldots & 0 & 0 & 0  \\
    0 & -h_{2} & \Lambda_{2} & -h'_{1} & \ldots & 0 & 0 & 0  \\
     &  &  & & \ddots    \\
    0 & 0 & 0 & 0 & \ldots & -h_{p-1} & \Lambda_{p-1} & -h'_{p-2}  \\
    -h'_{p-1} & 0 & 0 & 0 & \ldots & 0 & -h_{p} & \Lambda_{p}
\end{array} \right) = 0 \,,
\label{opendetform}
\ee
where 
\be 
h(u) &=& -\sinh^{2N}(u+\eta){\sinh(2u+2\eta)\over \sinh(2u+\eta)}
\sinh(u+\xi_{-}) \sinh(u-\xi_{+}) \,, \label{functionhopen} \\
h'(u) &=& h(-u-2\eta) \,, \label{functionhprime}
\ee
and $h_{k} = h(u + \eta k)$, $h'_{k} = h'(u + \eta k)$, 
$\Lambda_{k} = \Lambda(u + \eta k)$. We let
$(Q_{0} \,, Q_{1} \,, \ldots \,, Q_{p})$ be
the null vector of the matrix in (\ref{opendetform}); i.e.,
\be
\Lambda_{0} Q_{0} - h'_{-1} Q_{1} - h_{0} Q_{p} &=& 0 \,,  
\non \\
-h_{k} Q_{k-1} + \Lambda_{k} Q_{k} - h'_{k-1} Q_{k+1} &=& 0 \,, \qquad
k = 1\,, \ldots \,, p-1  \,, \non \\
-h'_{p-1} Q_{0} - h_{p} Q_{p-1} + \Lambda_{p} Q_{p}  &=& 0 \,.
\label{nullopen}
\ee
We make the Ansatz $Q_{k} = Q(u + \eta k)$, where
\be
Q(u) = \prod_{j=1}^{M} \sinh(u - u_{j}) \sinh(u + u_{j} + \eta) \,,
\label{openQ}
\ee 
which has the crossing symmetry $Q(u) = Q(-u-\eta)$. 
Eqs. (\ref{nullopen}) and (\ref{functionhprime}) imply
that the eigenvalues are given by
\be
\Lambda(u) = h(u) {Q(u-\eta)\over Q(u)} 
+ h(-u-\eta) {Q(u+\eta)\over Q(u)}  \,.
\label{openeigenvalues} 
\ee
We verify that this result is consistent with both
the periodicity (\ref{openperiodicity}) and crossing 
(\ref{transfercrossing}) properties of the transfer matrix.
The requirement that $\Lambda(u)$ be analytic at $u=u_{j}$ 
yields the Bethe Ansatz equations
\be
{h(u_{j})\over h(-u_{j}-\eta)} = 
-{Q(u_{j}+\eta)\over Q(u_{j}-\eta)} \,, \qquad j = 1 \,, \ldots \,, M \,.
\label{openBAeqs}
\ee

The results (\ref{functionhopen}), (\ref{openQ})-(\ref{openBAeqs}) for
the transfer-matrix eigenvalues and Bethe Ansatz equations agree with
those of Alcaraz {et al.} \cite{ABBBQ} and Sklyanin \cite{Sk}. 
Although we have assumed that $\eta$ has the values (\ref{etavalues}),
these results are true for general values of $\eta$.  As in the case
of the closed chain, one can establish that $M={N\over 2}-S^{z}$ by
considering the asymptotic behavior of $t(u)$ for $u \rightarrow
\infty$.

\subsubsection{The nondiagonal case $\kk_{+} = \kk_{-} \equiv \kk \,, \quad
\xi_{+} = \xi_{-} \equiv \xi \,, \quad N = \mbox{odd}$}

Finally, we consider the nondiagonal case $\kk_{+} = \kk_{-} \equiv
\kk \ne 0 \,, \quad \xi_{+} = \xi_{-} \equiv \xi \,, \quad N =
\mbox{odd}$.  For this case, the functional relations again have the
determinant representation (\ref{opendetform}), with
\be
h(u) = -\sinh^{2N}(u+\eta) {\sinh(2u+2\eta)\over \sinh(2u+\eta)}
\left( \sinh(u+\xi) \sinh(u-\xi) + \kk^{2} 
\sinh^{2}2u \right) \,. 
\label{functionhopen2} 
\ee 
It follows that the transfer-matrix eigenvalues and Bethe Ansatz
equations are again given by (\ref{openQ})-(\ref{openBAeqs}). 
However, unlike the two cases considered earlier which have a $U(1)$
symmetry, here the value of $M$ is fixed.  Indeed, the asymptotic
behavior (\ref{transfasympt}) implies that
\be
M={1\over 2}(N-1) \,.
\ee
We expect that, as in the previous cases, these results hold for
general values of $\eta$.

\section{Discussion}\label{sec:discuss}

We have seen that an approach used by Bazhanov and Reshetikhin
\cite{BR} to solve RSOS models, which is based on a model's functional
relations, is also an effective means of solving vertex models. 
Indeed, we have used this method to recover the known Bethe Ansatz
solutions of both the closed and open XXZ chains with $U(1)$ symmetry. 
Moreover, since this method does not rely on the existence of a
pseudovacuum state, we have also been able to use this method to solve
the special nondiagonal case (\ref{specialcase}) of the open chain.

Although we have focused here on vertex models associated with 
$sl_{2}$, it is clear that the same approach should be applicable to 
vertex models associated with higher-rank algebras.

Having found a model's functional relations, a crucial step in this
method is to reformulate the functional relations in determinant form. 
We have not yet succeeded to carry out this step for general values of
the boundary parameters of the open XXZ chain (\ref{Hamiltonian}).  It
would clearly be useful to find necessary and sufficient conditions
for the existence of a determinant representation of the functional
relations, as well as a systematic procedure for its construction.  We
hope to be able to report on these matters in a future publication.

\section*{Acknowledgments}

This work was supported in part by the National Science Foundation
under Grants PHY-9870101 and PHY-0098088.

\end{document}